\documentstyle[12pt]{article}
\input psfig

\newcommand{\beq}{\begin{equation}}
\newcommand{\eeq}{\end{equation}}
\newcommand{\bea}{\begin{eqnarray}}
\newcommand{\eea}{\end{eqnarray}}
\newcommand{\un}{\underline}
\newcommand{\half}{{\scriptstyle{{1\over 2}}}}

\newcommand{\real}{\relax{\rm I\kern-.18em R}}
\newcommand{\zahlen}{{\rm Z \!\! Z}}

\newcommand{\quat}{{\rm I \! H}}

\newcommand{\id}{\mbox{$id$}}

\newcommand{\Tr}{\mbox{\,Tr\,}}
\newcommand{\ad}{{\rm ad}}

\newcommand{\norm}[1]{\left\| #1 \right\|}
\newcommand{\sgbar}{\bar{\sg}}
\newcommand{\tr}{\mbox{\,tr\,}}
\newcommand{\al}{\alpha}
\newcommand{\Gm}{\Gamma}

\newcommand{\dl}{\delta}

\newcommand{\veps}{\varepsilon}

\newcommand{\lm}{\lambda}
\newcommand{\Lm}{\Lambda}
\newcommand{\sg}{\sigma}

\newcommand{\Om}{\Omega}
\newcommand{\Ss}[1]{\mbox{$\cal #1$}}
\newcommand{\pr}{\partial}

\newcommand{\Lkw}{\vec{L}_1^2}
\newcommand{\ip}[1]{\left\langle #1 \right \rangle }
\newcommand{\Order}[1]{\Ss{O}\left(#1\right)}

\begin{document}
\vskip-1cm
\hfill INLO-PUB-19/95
\vskip5mm
\begin{center}
{\LARGE{\bf{\underline{Global issues in gauge fixing}}}
\footnote{Talk at the ECT* workshop ``Non-perturbative
approaches to QCD'', July 10 - 29, 1995, Trento, Italy}
}\\
\vspace{1cm}
{\large Pierre van Baal
} \\
\vspace{1cm}
Instituut-Lorentz for Theoretical Physics,\\
University of Leiden, PO Box 9506,\\
NL-2300 RA Leiden, The Netherlands.\\
\end{center}
\vspace*{5mm}{\narrower\narrower{\noindent
\underline{Abstract:} We review the global issues associated
to gauge fixing ambiguities and their consequence for glueball spectroscopy.
To avoid infrared singularities the theory is formulated in a finite volume.
The examples of a cubic and spherical geometry will be discussed in some
detail. Our methods are not powerful enough to study the infinite volume
limit, but the results clearly indicate that for low-lying states,
wave functionals are sensitive to global gauge copies which we will argue
is equivalent to saying that they are sensitive to the geometric and
topological features of configuration space.}\par}

\section{Introduction}
Gauge theories with non-Abelian gauge groups lead not only to more complicated
potentials, but also to complicated kinetic terms in the Lagrangian or
Hamiltonian. The latter are a manifestation of the non-trivial Riemannian
geometry of the physical configuration space~\cite{bab}, formed by the set of
gauge orbits $\cal{A}/\cal{G}$ ($\cal{A}$ is the collection of connections,
$\cal{G}$ the group of local gauge transformations).
Most frequently, coordinates of this orbit space are chosen by
picking a representative gauge field on the orbit in a smooth and preferably
unique way. It is by now well known that linear gauge conditions like the
Landau or Coulomb gauge suffer from Gribov ambiguities~\cite{gri}. The reason
behind this is that topological obstructions prevent one from introducing
affine coordinates~\cite{sin} in a global way. In principle therefore, one
can introduce different coordinate patches with transition functions to
circumvent this problem~\cite{nah}. One way to make this specific is to base
the coordinate patches on the choice of a background gauge condition.
One could envisage to associate to each coordinate patch ghost fields
and extend BRST symmetry to include fields with non-trivial ``Grassmannian
sections'', although such a formulation is still in its infancy. Interesting
conjectures concerning non-perturbative spontaneous breakdown of BRST
invariance are implied by the work of Zwanziger and
collaborators~\cite{zwa,sch}, but will not be discussed here.

We will pursue, however, the issue of finding a fundamental domain for
non-Abelian gauge theories~\cite{sem} and its consequence for the glueball
spectrum in intermediate volumes. The finite volume context allows us to make
reliable statements on the non-perturbative contributions, because asymptotic
freedom guarantees that at small volumes the effective coupling constant is
small, such that high-momentum states can be treated perturbatively. Only
the lowest (typically zero or near-zero momentum) states will be affected by
non-perturbative corrections. We emphasize that it is essential that gauge
invariance is implemented properly at all stages. We will describe the results
mainly in the context of a Hamiltonian picture~\cite{chr} with wave functionals
on configuration space. Although rather cumbersome from a perturbative point
of view, where the covariant path integral approach of Feynman is vastly
superior, it provides more intuition on how to deal with non-perturbative
contributions to observables that do not vanish in perturbation theory.
An essential feature of the non-perturbative behaviour is that the wave
functional spreads out in configuration space to become sensitive to its
non-trivial geometry. If wave functionals are localized within regions
much smaller than the inverse curvature of the field space, the curvature
has no effect on the wave functionals.  At the other extreme, if the
configuration space has non-contractible circles, the wave functionals
are drastically affected by the geometry, or topology, when their support
extends over the entire circle. Instantons are of course the most important
examples of this. Not only the vacuum energy is affected by these
instantons, but also the low-lying glueball states and this is what we
are after to describe accurately, albeit in sufficiently small volumes.
The geometry of the finite volume, to be considered here, is the one
of a three-torus~\cite{lue,baa2} and a three-sphere~\cite{cut1,baa1}.

Some complications with so-called reducible connections~\cite{don}
(connections with a non-trivial stabilizer, i.e. subgroup of the gauge group
that leaves the connection invariant), discussed in detail in ref.~\cite{fuc},
will not occur in the sector where we study the dynamics. Although $A=0$ is
such a reducible connection, and would give rise to a curvature singularity in
configuration space, we know perfectly well how to deal with it by not fixing
the constant gauge transformations, which form the stabilizer of this vector
potential. Indeed the Coulomb gauge does not fix this gauge degree of freedom.
We simply demand that the wave functional is in the singlet representation
under the constant gauge transformations.

Finally we should mention that recently the issue of summing over all gauge
copies~\cite{hir} has been revived~\cite{lee} by studying a simple
soluble ``gauge'' model. It is shown in this model that singularities
which arise through the vanishing of the Faddeev-Popov determinant, which
is associated to the Gribov horizon, can be properly accounted for. It is
not clear that this can also be achieved in the case of Yang-Mills theories,
where things are far less tractable, although a recent paper by
Fujikawa~\cite{fuj} gives arguments in favour of this on the basis of a
careful BRST analysis. It is argued in ref.~\cite{lee} that selecting only
one gauge copy requires imposing artificial boundary conditions,
which changes the topology of the problem. This is exactly contrary to what
was this authors original motivation, namely that the non-trivial topology
enforces one to consider those boundary conditions~\cite{baa3}. In principle,
as we will see, these boundary conditions are well defined and only involve
identifications by a gauge transformation, under which we know exactly how
the wave functional transforms. Although it is correct to state that it is
difficult to develop the Feynman rules for perturbative calculations, it
should be emphasized that the global issues that arise from the gauge
invariance of the theory, leading to a topologically and geometrically
non-trivial configuration space, give rise to non-perturbative effects
that are not expected to be predicted reliably by perturbation theory.
The issue is to isolate the relevant non-perturbative contributions and
to include them in a way that will not violate the gauge invariance
of the theory. In that sense one can even take the extreme point of
view in ref.~\cite{joh} where it is attempted to eliminate all gauge
degrees of freedom by formulating the theory in terms of electric or
magnetic fields, although this leads to all sorts of technical difficulties,
but it has the advantage that approximations made after this reformulation
is implemented do not break the gauge invariance.

In the following we provide a quick review of the definition of the fundamental
domain. For gauge theories on a torus, when restricting oneself to the
zero-momentum modes, the fundamental domain is simple and leads to a reliable
way of computing the low-lying glueball spectrum in volumes up to half
a cubic fermi. We then describe the situation for the three sphere, where
also a rather complete picture of the fundamental domain for the low-lying
modes has been achieved.

\section{Gribov and fundamental regions}

An (almost) unique representative of the gauge orbit is found by
minimizing the $L^2$ norm of the vector potential along the gauge
orbit~\cite{sem,del1}
\beq
F_A(g)~\equiv\norm{^g A}^2~=~ -\int_M d^3x~
\tr \left( \left( g^{-1} A_i g + g^{-1} \pr_i g \right)^2\right),
\label{gAnorm}
\end{equation}
where the vector potential is taken anti-hermitian. For $SU(2)$,
in terms of the Pauli matrices $\tau_a$, one has:
\bea
A_i (x)&=& iA^a_i (x) \frac{\tau_a}{2}~,\nonumber\\
g (x) &=& \exp\left(X (x)\right),~X(x) = iX^a (x)
\frac{\tau_a}{2}.
\eea
Expanding around the minimum of eq.~(\ref{gAnorm}), one easily finds:
\bea
\norm{^g A}^2 &=& \norm{A}^2+2\int_M \tr(X
\partial_i A_i)+\int_M \tr (X^\dagger FP (A) X) \nonumber \\
&&+\frac{1}{3}\int_M\tr\left(X\left[[A_i,X],\partial_i X\right]\right)
+\frac{1}{12}\int_M\tr\left([D_iX,X][\partial_i X,X]\right)+\Order{X^5}.
\label{Xexpansie}
\eea
Where $FP(A)$ is the Faddeev-Popov operator $(\ad(A)X~\equiv~[A,X])$
\beq
FP (A)~=~-\partial_i D_i (A)~=~-\partial^2_i -\partial_i\ad(A_i).
\label{FPdef}
\eeq

At any local minimum the vector potential is therefore transverse,
$\partial_i A_i~=~0$, and $FP(A)$ is a positive operator. The
set of all these vector potentials is by definition the Gribov region
$\Omega$. Using the fact that $FP(A)$ is linear in $A$, $\Omega$ is
seen to be a convex subspace of the set of transverse connections $\Gamma$.
Its boundary $\partial \Omega$ is called the Gribov horizon. At the Gribov
horizon, the \underline{lowest} eigenvalue of the Faddeev-Popov operator
vanishes, and points on $\partial\Omega$ are hence associated with coordinate
singularities. Any point on $\partial\Omega$ can be seen to have a finite
distance to the origin of field space and in some cases even
uniform bounds can be derived~\cite{del2,zwa1}.

The Gribov region is the set of \underline{local} minima of the
norm functional (3) and needs to be further restricted to the
\un{absolute} minima to form a fundamental domain, which will be denoted
by $\Lambda$. The fundamental domain is clearly contained within the
Gribov region. To show that also $\Lambda$ is convex we note that
\bea
  \norm{^gA}^2 - \norm{A}^2 &=&
     \int \tr \left( A_i^2 \right)
     - \int \tr \left( \left( g^{-1} A_i g + g^{-1} \pr_i g \right)^2\right)
   \nonumber \\
  &=& \int \tr \left( g^\dagger FP_\half(A)~g \right)
       \equiv \ip{g,FP_\half(A)~g},
  \label{FPhalfdef}
\eea
where $FP_\half(A)$  is the Faddeev-Popov operator generalized to the
fundamental representation. Or for the gauge group SU(2) we have
\beq
  FP_t(A) = - \pr_i^2 - \frac{i}{t} A^a_i T_t^a \pr_i.
  \label{FPtdef}
\eeq
Here $\vec{T}_t$ are the hermitian gauge generators in the spin-$t$
representation:
\beq
  T^a_\half = \frac{\tau_a}{2} , \hspace{1.5cm}
  T^a_1 = \mbox{ad}(\frac{\tau_a}{2}).
\eeq
They are angular momentum operators that
satisfy $\vec{T}^2_t = t (t+1)\id$. At the critical points
$A \in \Gm$ of the norm functional,
(recall $\Gm = \{ A \in \Ss{A} | \pr_i A_i = 0 \}$),
$FP_t(A)$ is an hermitian operator. Furthermore, $FP_1(A)$ in that case
coincides with the Faddeev-Popov operator $FP(A)$ in eq.~(\ref{FPdef}).
We can define $\Lm$ in terms of the absolute minima (apart from the
boundary identifications) over $g \in \Ss{G}$ of
$\ip{g, FP_\half(A)~g}$
\beq
  \Lm =
  \{ A \in \Gm | \min_{g \in \Ss{G}} \ip{g, FP_\half(A)~g}=0 \}.
  \label{Lmdef}
\eeq
Using that $FP_\half(A)$ is linear in $A$ and assuming that $A^{(1)}$ and
$A^{(2)}$ are in $\Lm$ and therefore satisfy the equation
$\min_{g \in \Ss{G}} \ip{g, FP_\half(A)~g}=0$,
we find that $A=sA^{(1)}+(1-s)A^{(2)}$ satisfies the same identity
for all $s\in[0,1]$ (such that both $s$ and $(1-s)$ are positive). The line
connecting two points in $\Lm$, therefore lies within $\Lm$.

Its interior is devoid of gauge copies, whereas its boundary
$\partial\Lambda$ will in general contain gauge copies, which are
associated to those vector potentials where the absolute minima of
the norm functional are degenerate~\cite{baa3}. If this degeneracy is
continuous one necessarily has at least one zero eigenvalue for
$FP(A)$ and the Gribov horizon will touch the boundary of the fundamental
domain at these so-called singular boundary points. By singular we mean
here a coordinate singularity. We sketch the situation in figure 1.
\begin{figure*}{\tt}
\vspace{10.3cm}
\includegraphics{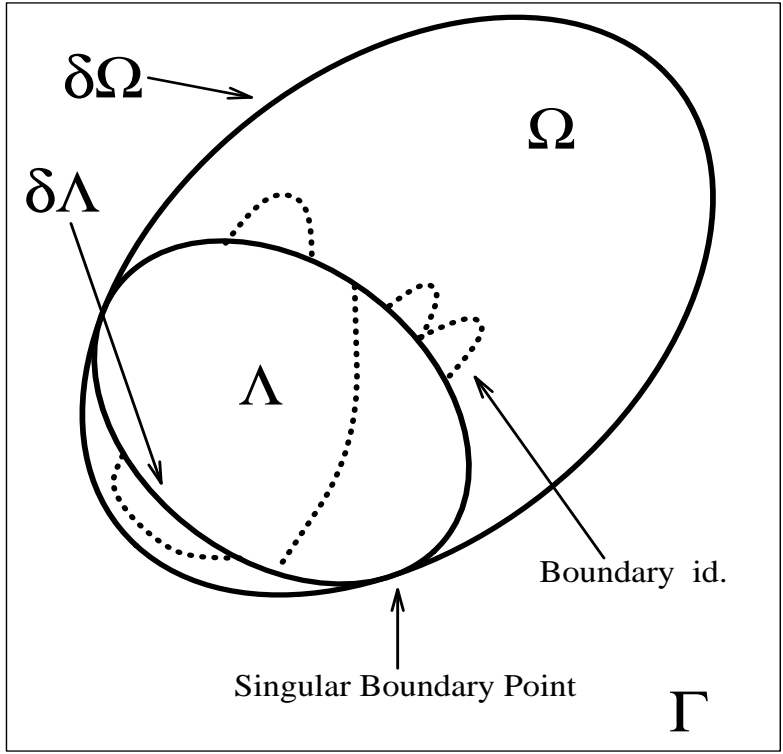}
\caption{Sketch of the Gribov and fundamental regions. The dotted lines
indicate the boundary identifications.}
\label{fig:fig1}
\end{figure*}
As mentioned already in the introduction, the constant gauge degree of freedom
is \underline{not} fixed by the Coulomb gauge condition and therefore one still
needs to divide by $G$ to get the proper identification
\beq
\Lm/G~=~ \cal{A}/\cal{G}.
\eeq

Here $\Lm$ is considered to be the set of absolute minima modulo
the boundary identifications, that remove the degenerate absolute minimum.
It is these boundary identifications that restore the non-trivial topology
of $\cal{A}/\cal{G}$. Furthermore, the
existence of non-contractible spheres~\cite{sin} allows one to argue that
singular boundary points are to be expected~\cite{baa3}. Consider the
intersection of a $d$-dimensional non-contractible sphere with $\Lm$.
The part in the interior of $\Lm$ is contractible and it can only become
non-contractible through the boundary identifications. It was falsely stated in
ref.~\cite{baa3} that this can only be the case if for the ($d$-1)-dimensional
intersection with the boundary, all points are to be identified.
As this would imply degeneracy of the the norm functional on a
($d$-1)-dimensional subspace, there would be at least $d$-1 zero-modes
for the Faddeev-Popov operator. Although each connected component of the
intersection with the interior of $\Lm$ is contractible, there can be more than
one connected component. In the case of two such components one can make
a non-contractible sphere by identifing points of the ($d$-1)-dimensional
boundary intersection of the first connected component with that of the
second and there is no necessity for a continuous degeneracy~\cite{tay}.

Not all singular boundary points, even those associated with continuous
degeneracies, need to be associated with non-contractible
spheres. Note that absolute minima of the norm functional are
degenerate along the constant gauge transformations, this is a
trivial degeneracy, also giving rise to trivial zero-modes for
the Faddeev-Popov operator, which we ignore. The action of $G$ is
essential to remove the curvature singularities mentioned above
and also greatly facilitates the standard Hamiltonian formulation
of the theory~\cite{chr}. There is no problem in dividing out $G$ by
demanding wave functionals to be gauge singlets (colourless states)
with respect to $G$. In practice this means effectively that one
minimizes the norm functional over $\Ss{G}/G$. When a singular boundary
point is not associated to a continuous degeneracy, the norm functional
undergoes a bifurcation, when we move from inside to outside the fundamental
(and Gribov) region. The stable absolute minimum turns into a saddle
point and two stable minima appear, as indicated in figure 2. These are
necessarily gauge copies of each other. The gauge transformation is
homotopically trivial as it reduces to the identity at the bifurcation
point, evolving continuously from there on. The explicit example of
ref.~\cite{baa3} is one where the connection is reducible and it
can be shown that the two stable minima that appear after the bifurcation
are gauge copies by constant gauge transformations~\cite{baat}, the
situation being subtle as the gauge transformation $g_1^{-1}g_2$
(see figure 2) is in this case not constant, but it is in the stabilizer
of the appropriate gauge field, up to a constant gauge transformation.
Examples of bifurcations at irreducible connections were explicitly found
for $S^3$, see ref.~\cite{heu} (app. A). We will come back to this.

\begin{figure*}{\tt}
\vspace{12.1cm}
\includegraphics{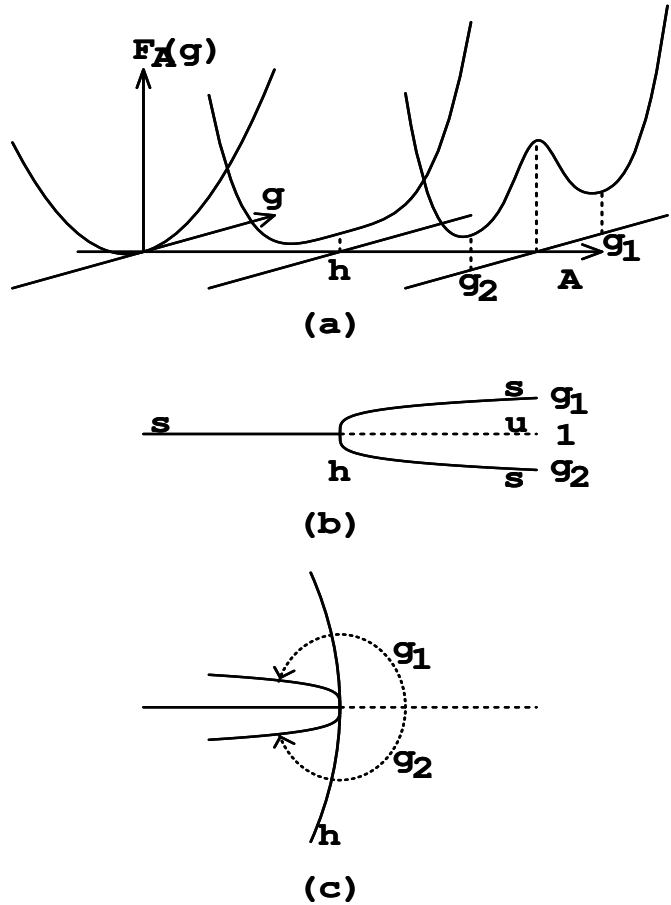}
\caption{Sketch of a singular boundary point due to a bifurcation of
the norm functional. It can be used to show that there are homotopically
trivial gauge copies inside the Gribov.}
\label{fig:fig2}
\end{figure*}

Also Gribov's original arguments for the existence of gauge copies~\cite{gri}
(showing that points just outside the horizon are gauge copies of points
just inside) can be easily understood from the perspective of bifurcations in
the norm functional. It describes the generic case where the zero-mode of
the Faddeev-Popov operator arises because of the coalescence of a stable
minimum with a saddle point with only one unstable direction, see
ref.~\cite{baa3} for more details and a discussion of the Morse theory
aspects that simplify the bifurcation analysis.

\section{Gauge fields on the three-torus}

Homotopical non-trivial gauge transformations are in one to one
correspondence with non-contractible loops in configuration space, which
give rise to conserved quantum numbers. The quantum numbers are like the
Bloch momenta in a periodic potential and label representations of the
homotopy group of gauge transformations. On the fundamental domain
the non-contractible loops arise through identifications of boundary points
(as will be demonstrated quite explicitly for the torus in the
zero-momentum sector). Although slightly more hidden, the fundamental
domain will therefore contain all the information relevant
for the topological quantum numbers. Sufficiently accurate knowledge of the
boundary identifications will allow for an efficient and natural
projection on the various superselection sectors (i.e. by choosing
the appropriate ``Bloch wave functionals''). All these features
were at the heart of the finite volume analysis on the torus~\cite{baa2} and
we see that they can in principle be naturally extended to the
full theory, thereby including the desired $\theta$ dependence. In the
next section this will be discussed in the context of the three-sphere.
In ref.~\cite{baa5} we proposed formulating the Hamiltonian theory on
coordinate patches, with homotopically non-trivial gauge transformations as
transition functions. Working with boundary conditions on the boundary
of the fundamental domain is easily seen to be equivalent and conceptually
much simpler to formulate. If there would be no singular boundary points
this would have provided a Hamiltonian formulation where all topologically
non-trivial information can be encoded in the boundary conditions. Still,
for the low-lying states in a finite volume, both on the three-torus and
the three-sphere, singular boundary points will not play an important role
in intermediate volumes.

Probably the most simple example to illustrate the relevance of the
fundamental domain is provided by gauge fields on the torus in the
abelian zero-momentum sector. For definiteness let us take $G=SU(2)$
and $A_i=i{C_i\over 2L}\tau_3$ ($L$ is the size of the torus). These
modes are dynamically motivated as they form the set of gauge
fields on which the classical potential vanishes. It is called the
vacuum valley (sometimes also referred to as toron valley) and one can
attempt to perform a Born-Oppenheimer-like approximation for deriving
an effective Hamiltonian in terms of these ``slow'' degrees of freedom.
To find the Gribov horizon, one easily verifies that the part of the spectrum
for $FP(A)$ that depends on $\vec C$, is given by $\lambda^{gh}_{\vec n}(\vec
C)=2\pi\vec n\cdot(2\pi\vec n\pm\vec C)$, with $\vec n\neq\vec 0$ an integer
vector. The lowest eigenvalue therefore vanishes if $C_k=\pm2\pi$. The Gribov
region is therefore a cube with sides of length $4\pi$, centred at the origin,
specified by $|C_k|\leq2\pi$ for all $k$, see figure 3.

The gauge transformation $g_{(k)}=\exp(\pi i x_k\tau_3/L)$ maps $C_k$ to
$C_k+2\pi$, leaving the other components of $\vec C$ untouched. As $g_{(k)}$
is anti-periodic it is homotopically non-trivial (they are 't Hooft's twisted
gauge transformations~\cite{tho2}). We thus see explicitly that gauge copies
occur inside $\Om$, but furthermore the naive vacuum $A=0$ has (many) gauge
copies under these shifts of $2\pi$ that lie on the Gribov horizon. It can
actually be shown for the Coulomb gauge that for any three-manifold, any
Gribov copy by a homotopically non-trivial gauge transformation of $A=0$ will
have vanishing Faddeev-Popov determinant~\cite{baa3}. Taking the symmetry
under homotopically non-trivial gauge transformations properly into account
is crucial for describing the non-perturbative dynamics and one sees that
the singularity of the Hamiltonian at Gribov copies of $A=0$, where the
wave functionals are in a sense maximal, could form a severe obstacle in
obtaining reliable results.

To find the boundary of the fundamental domain we note that the gauge copies
$\vec C=(\pi,C_2,C_3)$ and $\vec C=(-\pi,C_2,C_3)$ have equal norm. The
boundary of the fundamental domain, restricted to the vacuum valley formed by
the abelian zero-momentum gauge fields, therefore occurs where $|C_k|=\pi$,
well inside the Gribov region, see figure 3. The boundary identifications are
by the homotopically non-trivial gauge transformations $g_{(k)}$. The
fundamental domain, described by $|C_k|\leq \pi$, with all boundary points
regular, has the topology of a torus. To be more precise, as the remnant of
the constant gauge transformations (the Weyl group) changes $\vec C$ to
$-\vec C$, the fundamental domain $\Lm/G$ restricted to the abelian constant
modes is the orbifold $T^3/Z_2$. Generalizations to arbitrary gauge groups
were considered in ref.~\cite{baa5}. (The fundamental domain turns out to
coincide with the unit cell or ``minimal'' coordinate patch defined in
ref.~\cite{baa5}).

\begin{figure}{\tt}
\vspace{6.5cm}
\includegraphics{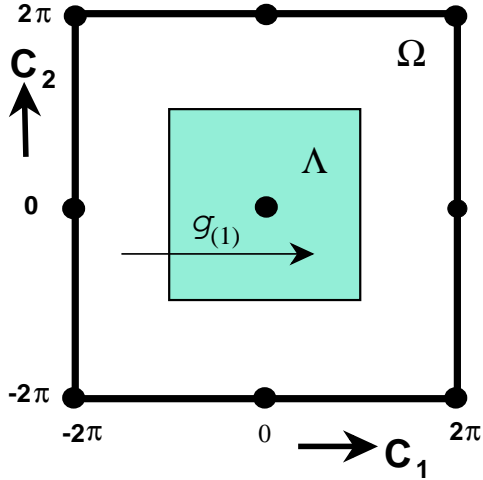}
\caption{A two dimensional slice of the vacuum valley along the $(C_1,C_2)$
plane. The fat square give the Gribov horizon, the grey square is the
fundamental domain. The dots at the Gribov horizon are Gribov copies
of the origin.}
\label{fig:fig3}
\end{figure}

Formulating the Hamiltonian on $\Lm$, with the boundary identifications
implied by the gauge transformations $g_{(k)}$, avoids the singularities
at the Gribov copies of $A=0$. ``Bloch momenta'' associated to the $2\pi$
shift, implemented by the non-trivial homotopy of $g_{(k)}$, label
`t Hooft's electric flux quantum numbers~\cite{tho2} $\Psi(C_k=-\pi)=\exp(
\pi ie_k)\Psi(C_k=\pi)$. Note that the phase factor is not arbitrary, but
$\pm 1$. This is because $g^2_{(k)}$ is homotopically trivial. In other
words, the homotopy group of these anti-periodic gauge transformations
is $Z_2^3$. Considering a slice of $\Lambda$ can obscure some of the
topological features. A loop that winds around the slice twice is
contractible in $\Lambda$ as soon as it is allowed to leave the slice.
Indeed including the lowest modes transverse to this slice will make the
$Z_2$ nature of the relevant homotopy group evident~\cite{baa2}. It should
be mentioned that for the torus in the presence of fields in the fundamental
representation (quarks), only periodic gauge transformations are allowed.
In that case it is easily seen that the intersection of the fundamental
domain with the constant abelian gauge fields is given by the domain
$|C_k|\leq 2\pi$, whose boundary coincides with the Gribov horizon. It is
interesting to note that points on $\partial\Om$ form an explicit example of
a continuous degeneracy due to a non-contractible sphere~\cite{zwa1}.

In weak coupling L\"uscher~\cite{lue} showed unambiguously that the
wave functionals are localized around $A=0$, that they are normalizable and
that the spectrum is discrete. In this limit the spectrum is insensitive to
the boundary identifications (giving rise to a degeneracy in the topological
quantum numbers). This is manifested by a vanishing electric flux energy,
defined by the difference in energy of a state with $|\vec e|=1$ and the
vacuum state with $\vec e=\vec 0$. Although there is no classical potential
barrier to achieve this suppression, it comes about by a quantum induced
barrier, in strength down by two powers of the coupling constant. This gives
a suppression~\cite{kol} with a factor $\exp(-S/g)$ instead of the usual factor
of $\exp(-8\pi^2/g^2)$ for instantons~\cite{tho}. Here $S=12.4637$ is the
action computed from the effective potential. At stronger coupling the wave
functional spreads out over the vacuum valley and the boundary conditions
drastically change the spectrum~\cite{baa2}. At this point the energy of
electric flux suddenly switches on.

The calculation is performed by integrating out the non-zero momentum degrees
of freedom, for which Bloch degenerate perturbation theory provides a rigorous
framework~\cite{blo,lue}, and gives an effective Hamiltonian. Near $A=0$, due
to the quartic nature of the potential energy $(F^a_{ij})^2$ for the
zero-momentum modes (the derivatives vanish and the field strength is
quadratic in the field), there is no separation in time scales between the
abelian and non-abelian modes. Away from $A=0$ one could further reduce the
dynamics to one along the vacuum valley, but near the origin this would be
a singular decomposition (the adiabatic approximation breaks down). However,
as long as the coupling constant is not too large, the wave functional can be
reduced to a wave function on the vacuum valley near $\partial\Lm$ where
the boundary conditions can be implemented. These boundary conditions are
formulated in a manner that preserves the invariance under constant
gauge transformation and the effective Hamiltonian is solved by Rayleigh-Ritz
(providing \un{also} lower bounds from the second moment of the Hamiltonian).
The influence of the boundary conditions on the low-lying glueball states
is felt as soon as the volume is bigger than an inverse scalar glueball
mass. We summarize below the ingredients that enter the calculations.

The effective Hamiltonian is expressed in terms of the coordinates $c_i^a$,
where $i=\{1,2,3\}$ is the spatial index ($c_0=0$) and $a=\{1,2,3\}$ is
the SU(2)-colour index. These coordinates are related to the zero-momentum
gauge fields through $A_i^a(x)=c_i^a/L$. We note that the field strength
is given by $ F_{ij}^a=-\varepsilon_{abd}c_i^bc_j^d/L^2$ and we introduce
the gauge-invariant ``radial'' coordinate $r_i=\sqrt{\sum_a c_i^ac_i^a}$.
The latter will play a crucial role in specifying the boundary
conditions. For dimensional reasons the effective Hamiltonian is proportional
to $1/L$. It will furthermore depend on $L$ through the renormalized coupling
constant ($g(L)$) at the scale $\mu=1/L$. To one-loop order one has (for small
$L$) $g(L)^2=12\pi^2/[-11\ln(\Lambda_{MS}L)]$. One expresses the masses and
the size of the finite volume in dimensionless quantities, like mass-ratios
and the parameter $z=mL$. In this way, the explicit dependence of $g$ on $L$
is irrelevant. This is also the preferred way of comparing results obtained
within different regularization schemes (i.e. dimensional and lattice
regularization). The effective Hamiltonian is now given by
\bea
L\cdot H_{eff}(c)&=&{g^2\over 2(1+\alpha_1 g^2)}\sum_{i,a}
{\partial^2\over \partial
{c_i^a}^2}+{1\over 4}({1\over g^2}+\alpha_2)\sum_{ij,a}{F_{ij}^a}^2\nonumber\\
&&+\gamma_1\sum_i r_i^2+\gamma_2\sum_i r_i^4+\gamma_3\sum_{i>j} r_i^2r_j^2+
\gamma_4\sum_i r_i^6+\gamma_5\sum_{i\neq j} r_i^2 r_j^4
+\gamma_6\prod_i r_i^2\nonumber\\
&&+\alpha_3\sum_{ijk,a}r_i^2{F_{jk}^a}^2+\alpha_4\sum_{ij,a}r_i^2{F_{ij}^a}^2
+\alpha_5{\det}^2c.
\eea

We have organized the terms according to the importance of their contributions,
ignoring terms quartic in the momenta. The first line gives (when ignoring
$\alpha_{1,2}$) the lowest order effective Hamiltonian, whose energy
eigenvalues are ${\cal O}(g^{2/3})$, as can be seen by rescaling $c$ with
$g^{2/3}$. Thus, in a perturbative expansion $c={\cal O}(g^{2/3})$. The
second line includes the vacuum-valley effective potential (i.e. the part
that does not vanish on the set of abelian configurations). These two lines
are sufficient to obtain the mass-ratios to an accuracy of better than
5\%. The third line gives terms of ${\cal O}(g^4)$ in the effective potential,
that vanish along the vacuum-valley. The coefficients (to two-loop order
for $\gamma_i$) are
\bea
\gamma_1=-3.0104661\cdot10^{-1}-(g/2\pi)^23.0104661\cdot10^{-1}&,&
\alpha_1=+2.1810429\cdot10^{-2},
\nonumber\\
\gamma_2=-1.4488847\cdot10^{-3}-(g/2\pi)^29.9096768\cdot10^{-3}&,&
\alpha_2=+7.5714590\cdot10^{-3},
\nonumber\\
\gamma_3=+1.2790086\cdot10^{-2}+(g/2\pi)^23.6765224\cdot10^{-2}&,&
\alpha_3=+1.1130266\cdot10^{-4},
\nonumber\\
\gamma_4=+4.9676959\cdot10^{-5}+(g/2\pi)^25.2925358\cdot10^{-5}&,&
\alpha_4=-2.1475176\cdot10^{-4},
\nonumber\\
\gamma_5=-5.5172502\cdot10^{-5}+(g/2\pi)^21.8496841\cdot10^{-4}&,&
\alpha_5=-1.2775652\cdot10^{-3},
\nonumber\\
\gamma_6=-1.2423581\cdot10^{-3}-(g/2\pi)^25.7110724\cdot10^{-3}&.&
\eea

The choice of boundary conditions, associated to each of the irreducible
representations of the cubic group $O(3,\zahlen)$ and to the states that
carry electric flux~\cite{tho2}, is best described by observing that the
cubic group is the semidirect product of the group of coordinate permutations
$S_3$ and the group of coordinate reflections $Z_2^3$. We denote the parity
under the coordinate reflection $c_i^a\rightarrow -c_i^a$ by $p_i=\pm 1$
($i$ fixed). The electric flux quantum number for the same direction will
be denoted by $q_i=\pm 1$.  This is related to the more usual additive
(mod 2) quantum number $e_i$ by $q_j=\exp(i\pi e_j)$. Note that for SU(2)
electric flux is invariant under coordinate reflections. If not all of the
electric fluxes are identical, the cubic group is broken to $S_2\times
Z_2^3$, where $S_2(\sim Z_2)$ corresponds to interchanging the two
directions with identical electric flux (unequal to the other electric flux).
If all the electric fluxes are equal, the wave functions are irreducible
representations of the cubic group. These are the four singlets $A_{1(2)}^\pm$,
which are completely (anti-)symmetric with respect to $S_3$ and have
each of the parities $p_i=\pm 1$. Then there are two doublets $E^\pm$,
also with each of the parities $p_i=\pm 1$ and finally one has four triplets
$T_{1(2)}^\pm$. Each of these triplet states can be decomposed into
eigenstates of the coordinate reflections.  Explicitly, for $T_{1(2)}^\pm$
we have one state that is (anti-)symmetric under interchanging the two-
and three-directions, with $p_2=p_3=-p_1=\mp 1$. The other two states are
obtained through cyclic permutation of the coordinates. Thus, any
eigenfunction of the effective Hamiltonian with specific electric flux
quantum numbers $q_i$ can be chosen to be an eigenstate of the parity
operators $p_i$. The boundary conditions of these eigenfunctions
$\Psi_{\vec q,\vec p}(c)$ are simply given by
\bea
\Psi_{\vec q,\vec p}(c)|_{_{r_i=\pi}}=0\quad,&\quad{\it if}\quad
p_iq_i=-1 \nonumber\\
{\partial\over\partial r_i}(r_i\Psi_{\vec q,\vec p}(c))|_{_{r_i=\pi}}=0\quad,
&\quad{\it if}\quad p_iq_i=+1
\eea
and one easily shows that with these boundary conditions the Hamiltonian is
hermitian with respect to the innerproduct $<\Psi,\Psi^\prime>=\int_{r_i
\leq\pi}d^9c\Psi^*(c)\Psi^\prime(c)$. For negative parity states ($\prod_i p_i
=-1$) this description is, however, not accurate~\cite{voh2} as parity
restricted to the vacuum valley is equivalent to a Weyl reflection
(a remnant of the invariance under constant gauge transformations).

\begin{figure}{\tt}
\vspace{7.5cm}
\includegraphics{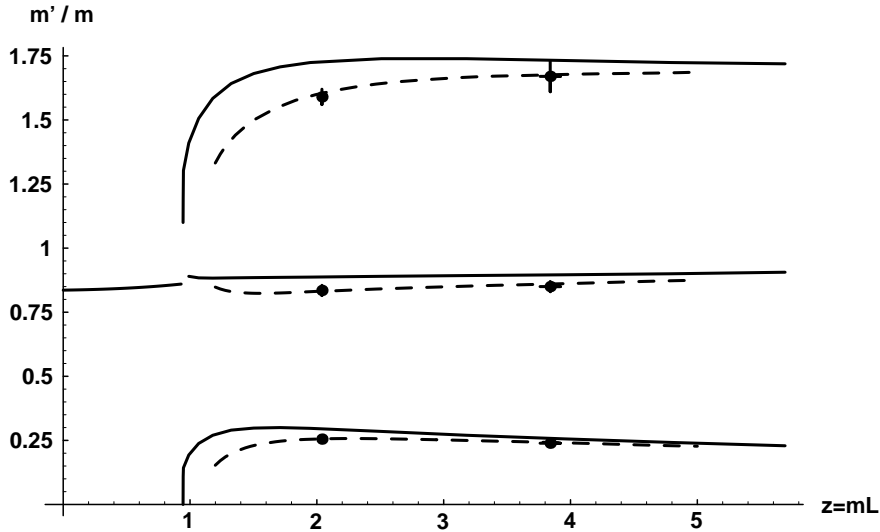}
\caption{
Mass ratios $m^\prime/m$ as a function of $L$ in units of the inverse
scalar glueball mass $m^{-1}$, $z=mL$. The analytic results are given
by the full (continuum) and dashed ($4^3$ lattice) curves. Represented are
the the square root of the energy of electric flux per unit length (bottom),
the $E$ tensor mass (middle) and the $T_2$ tensor mass (top). The latter
two are (almost) degenerate below $z=0.95$, which is also where the electric
flux energy is exponentially suppressed. Where no error bars on the Monte
Carlo data are visible they are smaller than the size of the data points.
}
\label{fig:fig4}
\end{figure}

After correcting for lattice artefacts~\cite{lat}, the (semi-)analytic results
agree extremely well with the best lattice data~\cite{tep} (with statistical
errors of 2\% to 3\%) up to a volume of about .75 fermi, or about five times
the inverse scalar glueball mass. In figure 4 we present the comparison for a
lattice of spatial size $4^3$. Monte Carlo data~\cite{tep} are most accurate
for this lattice size. For more detailed comparisons see ref.~\cite{lat}.
The analytic results below $z=0.95$ are due to L\"uscher and
M\"unster~\cite{lue}, which is where the spectrum is insensitive to the
identifications at the boundary of $\Lm$. Apart from the corrections for the
lattice artefacts, generalization to SU(3) was established by
Vohwinkel~\cite{voh}, with qualitatively similar results. In large volumes
the rotational symmetry should be restored, as is observed from lattice
simulations. Most conspicuously the tensor state in finite volumes is split
in a doublet $E$, with a mass that is roughly 0.9 times the scalar $A_1$ mass
and a triplet $T_2$ with a mass of roughly 1.7 times the scalar mass. Note
that the multiplicity weighted average is approximately 1.4 times the scalar
mass, agreeing well with what was found at large volumes from lattice
data~\cite{tep}. There has been similar studies using spatial twisted
boundary conditions (with differing behaviour of the tensor and electric
flux states in intermediate volumes), see ref.~\cite{ste}.

\begin{figure}{\tt}
\vspace{6cm}
\includegraphics{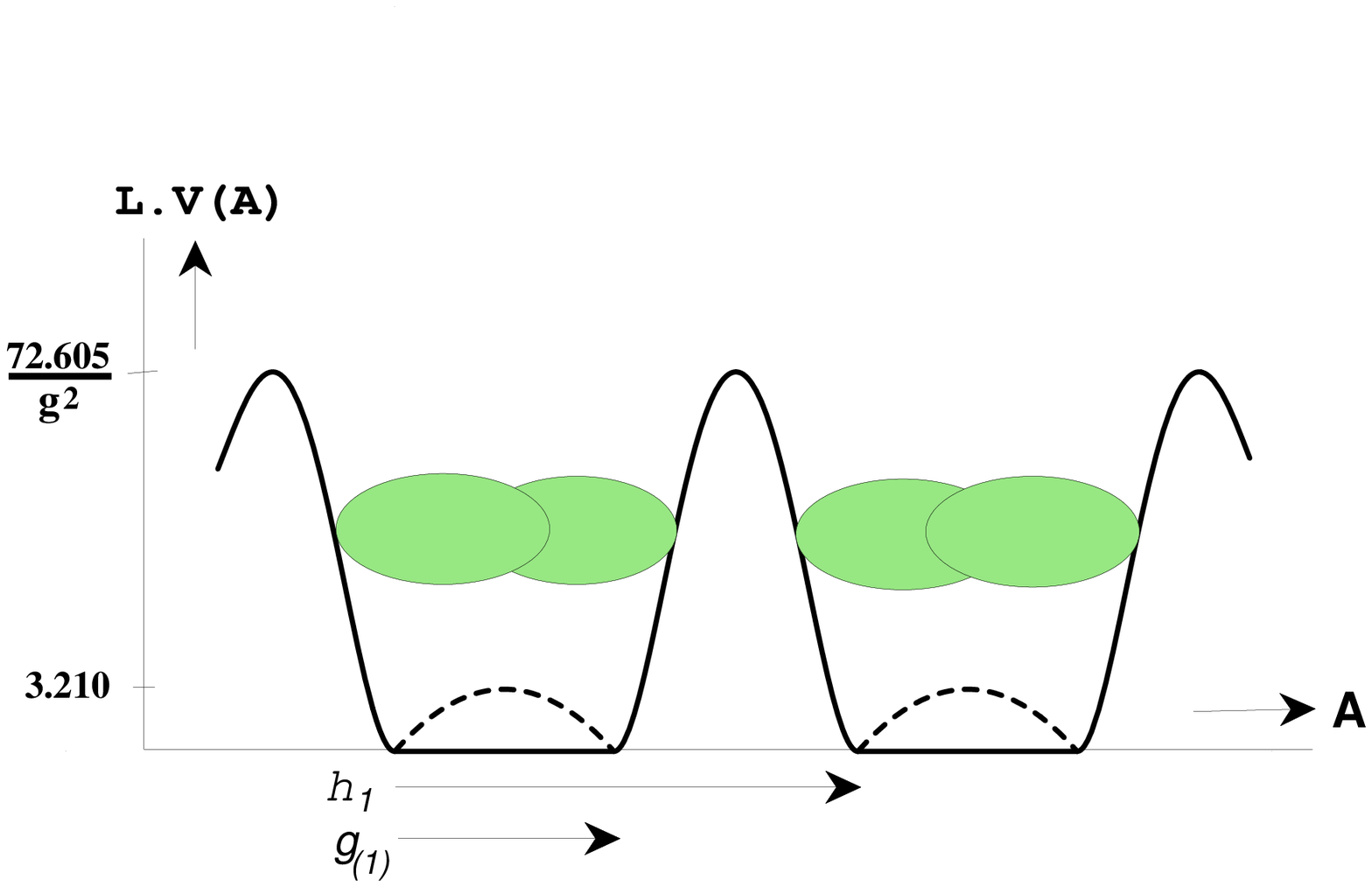}
\caption{``Artistic impression'' of the potential for the three-torus. Shown
are two vacuum valleys, related to each other by a gauge transformation $h_1$
with winding number 1, whose shape in the direction perpendicular to the
valley depends on the position along the valley. The induced one-loop effective
potential, of height $3.210/L$, has degenerate minima related to each other
by the anti-periodic gauge transformations $g_{(k)}$. The classical barrier,
separating the two valleys, has the height $72.605/Lg^2$.
}
\label{fig:fig5}
\end{figure}

At large volumes extra degrees of freedom start to behave non-perturbatively.
To demonstrate this, the minimal barrier height that separates two vacuum
valleys that are related by gauge transformations with non-trivial winding
number
\beq
\nu(g)={1\over 24\pi^2}\int_M\Tr((g^{-1}dg)^3),
\eeq
was found to be $72.605/Lg^2$, using the lattice approximation and
carefully taking the continuum limit~\cite{mar}. The situation is sketched in
figure 5. One can now easily find for which volume the energy of the level
that determines the glueball mass (defined by the difference with the
groundstate energy) starts to be of the order of this barrier height.
This turns out to be the case for $L$ roughly 5 to 6 times the correlation
length set by the scalar glueball mass. We expect, as will be shown for the
three-sphere, that the boundary of the fundamental domain along the path
in field space across the barrier (which corresponds to the instanton path
if we parametrize this path by Euclidean time $t$), occurs at the saddle
point (which we call a finite volume sphaleron) in between the two minima.
The degrees of freedom along this tunnelling path go outside of the
space of zero-momentum gauge fields and if the energy of a state flows
over the barrier, its wave functional will no longer be exponentially
suppressed below the barrier and will in particular be non-negligible at
the boundary of the fundamental domain. Boundary identifications in
this direction of field space now become dynamically important too.
The relevant ``Bloch momentum'' is in this case obviously the
$\theta$ parameter, as wave functionals pick up a phase factor $e^{i\theta}$
under a gauge transformation with winding number one. For many of the
intricacies in describing instantons on a torus we refer to ref.~\cite{mar2}.
On the three-torus we have therefore achieved a self-contained picture
of the low-lying glueball spectrum in intermediate volumes from first
principles with \un{no free parameters}, apart from the overall scale.

\section{Gauge fields on the three-sphere}

The reason to consider the three-sphere lies in the fact that the conformal
equivalence of $S^3\times\real$ to $\real^4$ allows one to construct
instantons explicitly~\cite{hos,baa1,smi}. This greatly simplifies the study
of how to formulate $\theta$ dependence in terms of boundary conditions on
the fundamental domain, and indeed we will see that for $S^3$ simple enough
results can be obtained to address this question~\cite{heu,heu2}.
The disadvantage of the three-sphere is that in large volumes the corrections
to the glueball masses are no longer exponential~\cite{lue}. In this
respect the use of twisted boundary conditions~\cite{ste} offers a viable
alternative, but only numerical solutions for the relevant instantons
are know~\cite{mar3}.

We will summarize the formalism that was developed in~\cite{baa1}.
Alternative formulations, useful for diagonalizing the Faddeev-Popov and
fluctuation operators, were given in ref.~\cite{cut1}, whereas for
the explicit formulation of instantons, ref.~\cite{smi} introduces
stereographic coordinates (demonstrating that simplicity is in the eye of
the beholder). We embed $S^3$ in $\real^4$ by considering the unit sphere
parametrized by a unit vector $n_\mu$. It is particularly useful to introduce
the unit quaternions $\sg_\mu$ and their conjugates $\bar{\sg}_\mu =
\sg^\dagger_\mu$ by
\beq
  \sg_\mu = ( \id ,i \vec{\tau}), \hspace{1.5cm}
  \bar{\sg}_\mu = ( \id,- i \vec{\tau}).
\eeq
They satisfy the multiplication rules
\beq
 \sg_\mu \sgbar_\nu = \eta^\al_{\mu \nu} \sg_\al, \hspace{1.5cm}
  \sgbar_\mu \sg_\nu = \bar{\eta}^\al_{\mu \nu} \sg_\al,
\eeq
where we used the 't Hooft $\eta$ symbols~\cite{tho}, generalized slightly to
include a component symmetric in $\mu$ and $\nu$ for $\al=0$. We can use
$\eta$ and $\bar{\eta}$ to define orthonormal framings~\cite{lue2} of $S^3$,
which were motivated by the particularly simple form of the instanton
vector potentials in these framings. The framing for $S^3$ is
obtained from the framing of $\real^4$ by restricting in the following
equation the four-index $\al$ to a three-index $a$
(for $\al = 0$ one obtains the normal on $S^3$):
\beq
  e^\al_\mu = \eta^\al_{\mu \nu} n_\nu , \hspace{1.5cm}
  \bar{e}^\al_\mu = \bar{\eta}^\al_{\mu \nu} n_\nu.
\eeq
Note that $e$ and $\bar{e}$ have opposite orientations. Each framing defines
a differential operator and associated (mutually commuting) angular
momentum operators $\vec{L}_1$ and $\vec{L}_2$:
\beq
  \pr^i = e^i_\mu \frac{\pr}{\pr x^\mu},\quad L_1^i = \frac{i}{2}~\pr^i\quad,
  \quad\bar{\pr}^i = \bar{e}^i_\mu \frac{\pr}{\pr x^\mu},\quad
  L_2^i = \frac{i}{2}~\bar{\pr}^i.
\eeq
It is easily seen that $\Lkw = \vec{L}_2^2$, which has eigenvalues
$l(l+1)$, with $l=0,\half,1,\cdots$.

The (anti-)instantons~\cite{bel} in these framings, obtained from those
on $\real^4$ by interpreting the radius in $\real^4$ as the
exponential of the time $t$ in the geometry $S^3 \times \real$, become
\beq
  A_0 = \frac{\vec{\veps} \cdot \vec{\sg}}{2 ( 1 + \veps \cdot n )}
  , \hspace{1.5cm}
  \vec A = \frac{\vec{\sg} \wedge \vec{\veps} -( u + \veps \cdot n )
  \vec{\sg}} {2 ( 1 + \veps \cdot n )},\label{vecA}
\eeq
where
\beq
   u = \frac{ 2 s^2}{1 + b^2 + s^2} , \hspace{1.5cm}
   \veps _\mu = \frac{2 s b_\mu}{1 + b^2 + s^2} , \hspace{1.5cm}
   s = \lm e^t.
\eeq
Here $\vec\veps$ and $\vec A$ are defined with respect to the
framing $e^a_\mu$ for instantons and with respect to the framing
$\bar{e}^a_\mu$ for anti-instantons.
The instanton describes tunnelling from $A = 0$ at $t = - \infty$ to
$A_a = - \sg_a$ at $t = \infty$, over a potential barrier at $t=0$
that is lowest when $b_\mu \equiv 0$. This configuration corresponds to a
sphaleron~\cite{kli}, i.e.\ the vector potential $A_a = -\half\sg_a$
is a saddle point of the energy functional with one unstable mode,
corresponding to the direction ($u$) of tunnelling. At $t = \infty$,
$A_a = - \sg_a$ has zero energy and is a gauge copy of $A_a = 0$ by a
gauge transformation $g = n \cdot \sgbar$ with winding number one.

We will be concentrating our attention to the modes that are degenerate
in energy to lowest order with the modes that describe tunnelling through
the sphaleron and "anti-sphaleron". The latter is a gauge copy by a gauge
transformation $g = n \cdot \sg$ with winding number $-1$ of the sphaleron.
The two dimensional space containing the tunnelling paths through these
sphalerons is consequently parametrized by $u$ and $v$ through
\beq
  A_\mu(u,v)=\left(-u e^a_\mu-v\bar{e}^a_\mu \right)\frac{\sg_a}{2}.
\eeq
The gauge transformation with winding number $-1$ is easily seen to
map $(u,v) = (w,0)$ into $(u,v) = (0,2-w)$.
The 18 dimensional space is defined by
\beq
  A_\mu(c,d)=\left(c^a_i  e^i_\mu+d^a_j\bar{e}^j_\mu \right)
  \frac{\sg_a}{2}=A_i(c,d)e_\mu^i.
  \label{Acddef}
\eeq
The $c$ and $d$ modes are mutually orthogonal and satisfy
the Coulomb gauge condition:
\beq
  \pr_i A_i(c,d) = 0.
\eeq
This space contains the $(u,v)$ plane through $c^a_i = -u \dl^a_i$ and
$d^a_i = -v \dl^a_i$. The significance of this 18 dimensional space is that the
energy functional~\cite{baa1}
\beq
  \Ss{V}(c,d) \equiv - \int_{S^3} \frac{1}{2} \tr(F_{ij}^2)
  = \Ss{V}(c) + \Ss{V}(d) + \frac{2 \pi^2}{3}
   \left\{ (c^a_i)^2 (d^b_j)^2 - (c^a_i d^a_j)^2 \right\}\label{pot}
,\eeq
\beq
  \Ss{V}(c) = 2 \pi^2 \left\{ 2 (c^a_i)^2 + 6 \det c +
  \frac{1}{4}[(c^a_i c^a_i)^2 - (c^a_i c^a_j)^2 ] \right\} ,
\eeq
is degenerate to second order in $c$ and $d$. Indeed, the quadratic
fluctuation operator \Ss{M} in the Coulomb gauge, defined by
\bea
  - \int_{S^3} \frac{1}{2} \tr(F_{ij}^2)
  &=& \int_{S^3} \tr(A_i \Ss{M}_{ij} A_j) + \Order{A^3},\nonumber \\
  \Ss{M}_{ij} &=& 2\vec{L}_1^2\dl_{ij} + 2 \left( \vec{L}_1 + \vec{S}
  \right)^2_{ij},\quad S^a_{ij}=-i \veps_{aij},\label{fluct}
\eea
has $A(c,d)$ as its eigenspace for the (lowest) eigenvalue $4$. These
modes are consequently the equivalent of the zero-momentum modes on the
torus, with the difference that their zero-point frequency does not vanish.
An effective Hamiltonian for the $c$ and $d$ modes is here derived from
the one-loop effective action and errors due to an adiabatic approximation
are not necessarily suppressed by powers of the coupling constant.
Nevertheless, one expects to achieve an approximate understanding of
the non-perturbative dynamics in this way~\cite{heu2}.

\begin{figure*}{\tt}
\vspace{6.7cm}
\includegraphics{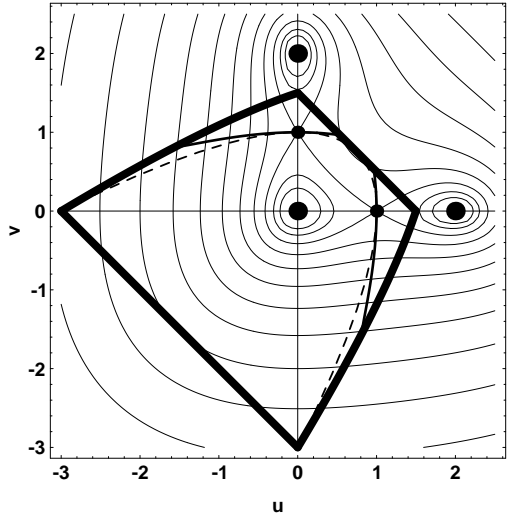}
\caption{
Location of the classical vacua (large dots), sphalerons (smaller
dots), the Gribov horizon (fat sections), the boundary of $\tilde\Lm$
(dashed curves) and part of the boundary of the fundamental domain (full
curves). Also indicated are the lines of equal potential in units of
$2^n$ times the sphaleron energy.
.}
\label{fig:fig6}
\end{figure*}

$FP_\half(A)$ in eq.~(\ref{FPhalfdef}) is defined as an hermitian
operator acting on the vector space $\Ss{L}$ of functions $g$ over $S^3$
with values in the space of the quaternions $\quat=\{q_\mu\sg_\mu|
q_\mu\in\real\}$. The gauge group $\Ss{G}$ is contained in $\Ss{L}$
by restricting to the unit quaternions:
$\Ss{G}=\{g\in\Ss{L}|g=g_\mu\sg_\mu,g_\mu\in\real,g_\mu g_\mu = 1 \}$.
For arbitrary gauge groups $\Ss{L}$ is defined as the algebra generated
by the identity and the (anti-hermitian) generators of the algebra.
When minimizing the same functional over the larger space $\Ss{L}$
one obviously should find a smaller space $\tilde{\Lm} \subset \Lm$.
Since $\Ss{L}$ is a linear space $\tilde\Lm$ can also be specified by
the condition that $FP_\half(A)$ be positive,
\beq
  \tilde{\Lm} =
  \{ A \in \Gm | \ip{g, FP_\half(A)~g}\geq0,\ \forall g\in\Ss{L} \}.
  \label{Lmtildef}
\eeq
As for the Gribov horizon, the boundary of $\tilde\Lm$ is therefore determined
by the location where the lowest eigenvalue vanishes. For the $(c,d)$ space
it can be shown~\cite{heu} that the boundary $\partial\tilde{\Lm}$ will touch
the Gribov horizon $\partial\Om$. This establishes the
existence of singular points on the boundary of the fundamental domain due
to the inclusion $\tilde{\Lm} \subset \Lm \subset \Om$.
By showing that the fourth order term in eq.~(\ref{Xexpansie}) is positive
(see app.~A of ref.~\cite{heu}) this is seen to correspond to the situation
as sketched in figure 1.

One can make convenient use of the ${\rm SU}(2)^3$ symmetry generated
by $\vec L_1$, $\vec L_2$ and $\vec T$ to calculate explicitly the spectrum
of $FP_t(A)$. One has
\beq
  FP_t(A(c,d)) = 4 \Lkw - \frac{2}{t} c^a_i T_t^a L_1^i
   - \frac{2}{t} d^a_i T_t^a L_2^i,
\eeq
which commutes with $\Lkw = \vec{L}_2^2$, but for arbitrary $(c,d)$
there are in general no other commuting operators (except for
a charge conjugation symmetry when $t=\half$). Restricting to
the $(u,v)$ plane one easily finds that
\beq
  FP_t(A(u,v)) = 4 \Lkw + \frac{2}{t} u \vec{L}_1 \cdot \vec{T}_t
    + \frac{2}{t} v \vec{L}_2 \cdot \vec{T}_t,
    \label{FPuvdef}
\eeq
which also commutes with the total angular momentum $\vec{J}_t=\vec{L}_1
+\vec{L}_2+\vec{T}_t$ and is easily diagonalized. Figure 6 summarizes
the results for this $(u,v)$ plane and also shows the equal-potential
lines as well as exhibiting the multiple vacua and the sphalerons. As it
is easily seen that the two sphalerons are gauge copies (by a unit winding
number gauge transformation) with equal norm, they lie on $\partial\Lm$,
which can be extend by perturbing around these sphalerons~\cite{baa4}.

\begin{figure*}{\tt}
\vspace{5.6cm}
\includegraphics{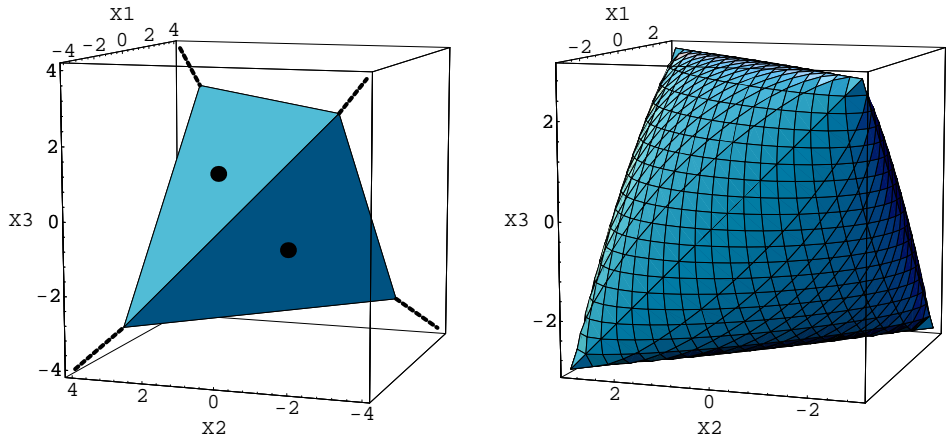}
\caption{
The fundamental domain (left) for constant gauge fields on $S^3$, with respect
to the ``instanton'' framing $e_\mu^a$, in the diagonal representation
$A_a=x_a\sg_a$ (no sum over $a$). By the dots on the faces we indicate the
sphalerons, whereas the dashed lines represent the symmetry axes of the
tetrahedron. To the right we display the Gribov horizon, which encloses
the fundamental domain, coinciding with it at the singular boundary points
along the edges of the tetrahedron.  }
\label{fig:fig7}
\end{figure*}

To obtain the result for general $(c,d)$ one can use the
invariance under rotations generated by $\vec{L}_1$ and $\vec{L}_2$
and under constant gauge transformations generated by $\vec{T}_t$,
to bring $c$ and $d$ to a standard form, or express
$\det\left(FP_t(A(c,d))|_{l = \half}\right)$, which determines the
locations of $\partial\Om$ and $\partial\tilde\Lm$, in terms of invariants.
We define the matrices $X$ and $Y$ by $X^a_b = (c c^t)^a_b$ and
$Y^a_b = (d d^t)^a_b$, which allows us to find
\bea
\det\left(FP_\half (A(c,d))|_{l = \half}\right)&=&
[81 - 18 \Tr (X +Y) + 24 ( \det c + \det d)\nonumber\\
 & & - (\Tr(X - Y))^2 + 2 \Tr((X-Y)^2)]^2.
  \label{halfcd}
\eea
The two-fold multiplicity is due to charge conjugation symmetry.
The expression for $t=1$, that determines the location of the
Gribov horizon in the $(c,d)$ space, is given in app.~B of ref.~\cite{heu}.
If we restrict to $d=0$ the result simplifies considerably. In that case
one can bring $c$ to a diagonal form $c_i^a=x_i\delta_i^a$. The rotational
and gauge symmetry reduce to permutations of the $x_i$ and simultaneous
changes of the sign of two of the $x_i$. One easily finds the invariant
expression ($\Tr(X)=\sum_i x_i^2$ and $\det c=\prod_i x_i$)
\beq
  \det\left(FP_1(A(c,0))|_{l = \half}\right)
= \left( 2 \det c - 3 \Tr(X) + 27 \right)^4.
\label{deteend0}
\eeq

In figure 7 we present the results for $\Lm$ and $\Om$. In this particular
case, where $d=0$, $\Lambda$ coincides with $\tilde\Lambda$, a consequence of
the convexity and the fact that both the sphalerons (indicated by the dots)
and the edges of the tetrahedron lie on $\partial\Lambda$, the latter also
lying on $\partial\Om$. It is essential that the sphalerons do not lie on
the Gribov horizon and that the potential energy near $\partial\Om$ is
relatively high. This is why we can take the boundary identifications near
the sphalerons into account without having to worry about singular boundary
points, as long as the energies of the low-lying states will be not much
higher than the energy of the sphaleron. It allows one to study the
glueball spectrum as a function of the CP violating angle $\theta$, but
more importantly it incorporates for $\theta=0$ the noticeable influence
of the barrier crossings, i.e. of the instantons. For details see~\cite{heu2}.

\section{Discussion}
We have analysed in detail the boundary of the fundamental
domain for SU(2) gauge theories on the three-torus and three-sphere.
It is important to note that it is necessary to divide $\Ss{A}$ by the set
of \un{all} gauge transformations, including those that are homotopically
non-trivial, to get the physical configuration space. All the non-trivial
topology is then retrieved by the identifications of points on the boundary
of the fundamental domain.

As we already mentioned in the introduction, the knowledge of the
boundary identifications is important in the case that the wave functionals
spread out in configuration space to such an extent that they become
sensitive to these identifications. This happens at large volumes, whereas
at very small volumes the wave functional is localized around $A=0$ and
one need not worry about these non-perturbative effects. That these effects
can be dramatic, even at relatively small volumes (above a tenth of a fermi
across), was demonstrated for the case of the torus.
However, for that case the structure of the fundamental domain (restricted
to the abelian zero-energy modes) is a hypercube and deviates
considerably from the fundamental domain of the three-sphere. Results for
the spectrum in the latter case recently became available~\cite{heu2} and
indicate that the tensor to scalar glueball mass ratio is compatible
in volumes around one fermi. For the three-sphere the tensor glueball
is of course not split into a doublet and triplet representation.

It should be noted that the shape of $\Lm$ is independent of $L$ if the gauge
field is expressed in units of $1/L$. Suppose that the coupling constant
will grow without bound. This would make the potential irrelevant and
makes the wave functional spread out over the whole of field space
(which could be seen as a strong coupling expansion). If the kinetic
term would have been trivial the wave functionals would be ``plane waves''
on a space with complicated boundary conditions. In that case it seems
unavoidable that the infinite volume limit would depend on the geometry
(like $T^3$ or $S^3$) that is scaled-up to infinity. With the non-triviality
of the kinetic term this conclusion cannot be readily made and our present
understanding only allows comparison in volumes around one cubic fermi.
However, one way to avoid this undesirable dependence on the geometry
is that the vacuum is unstable against domain formation. As periodic
subdivisions are space filling on a torus, this seems to be the preferred
geometry to study domain formation. In a naive way it will give the
correct string tension (flux conservation tells us to ``string'' the
domains that carry electric flux) and tensor to scalar mass ratio
(averaging over the orientations of the domains is expected to lead to
a multiplicity weighted average of the $T_2$ and $E$ masses).
Furthermore, the natural dislocations of such a domain picture are gauge
dislocations. The point-like gauge dislocations in four dimensions are
instantons and in three dimensions they are monopoles. Their density is
expected to be given roughly as one per domain (with a volume of around
0.5 cubic fermi). Also the coupling constant will stop running at the scale
of the domain size. We have discussed this elsewhere and refer the reader
to refs.~\cite{baa5,baa6} for further details, as the ideas in this
direction remain speculative. In the context of twisted boundary
conditions, related ideas were recently developed in ref.~\cite{gon2}.

\section*{Acknowledgements}

I wish to thank Mitya Diakonov for inviting me to this workshop and
the staff of the ECT* for creating such a nice environment. The lively
``Russian'' style of the workshop was a delight and I am grateful for
discussions with many of the participants.
In particular I thank Emil Akhmedov, Maxim Chernodub, Mitya Diakonov,
Adriano DiGiacomo, Tony Gonzalez-Arroyo, Victor Petrov, Misha Polikarpov,
Tsuneo Suzuki, Mike Teper and Jac Verbaarschot.
I also thank Washington Taylor for pointing out an error in the topological
argument for the existence of singular boundary points.

\end{document}